\documentclass{elsart}

 \usepackage{graphicx}
\usepackage{multirow}

\newcommand{\lsim}{~{\buildrel < \over {_\sim}}~}
\newcommand{\gsim}{~{\buildrel > \over {_\sim}}~}
\newcommand{\sqrtsNN}{\sqrt{s_{\scriptscriptstyle{{\rm NN}}}}}
\newcommand{\av}[1]{\left\langle #1 \right\rangle}

\newcommand{\gev}{\mathrm{GeV}}
\newcommand{\tev}{\mathrm{TeV}}
\newcommand{\fm}{\mathrm{fm}}

\newcommand{\RAA}{R_{\rm AA}}
\newcommand{\RCP}{R_{\rm CP}}
\newcommand{\pt}{p_{\rm t}}
\renewcommand{\d}{{\rm d}}

\newcommand {\apgt} {\ {\raise-.5ex\hbox{$\buildrel>\over\sim$}}\ }
\newcommand {\aplt} {\ {\raise-.5ex\hbox{$\buildrel<\over\sim$}}\ }

\begin{document}

\begin{frontmatter}



\title{Effect of heavy-quark energy loss on the  muon differential
production cross section \mbox{in {\rm Pb--Pb} collisions at $\sqrtsNN=5.5~\tev$}}

\author[a]{Z. Conesa del Valle\thanksref{permaddr}},
\author[b]{A. Dainese},
\author[c]{H.-T. Ding},
\author[a]{G. Mart\'inez Garc\'ia},
and~\author[c]{D.C. Zhou}

\thanks[permaddr]{Now at {\it LLR (CNRS/IN2P3 - Ecole Polytechnique) Palaiseau, France}}

\address[a]{\scriptsize Subatech (CNRS/IN2P3 - Ecole des Mines - Universit\'e
de Nantes) Nantes, France}
\address[b]{\scriptsize INFN - Laboratori Nazionali di
Legnaro, 35020 Legnaro (Padova), Italy}
\address[c]{\scriptsize Institute of Particle Physics, Central China Normal
University, Wuhan 430079, China}

\begin{abstract}
We study the nuclear modification factors $\RAA$ and $\RCP$
of the high transverse momentum ($5<p_{\rm t}<60~\gev/c$)
distribution of muons
in \mbox{Pb--Pb} collisions at LHC energies. We consider two
pseudo-rapidity ranges covered by the LHC experiments: $|\eta|<2.5$ and
$2.5<\eta<4$.
Muons from semi-leptonic decays of heavy quarks (c and b) and
from leptonic decays of weak gauge bosons (W and Z)
are the main contributions to
the muon $\pt$ distribution above a few $\gev/c$.
We compute the heavy quark contributions using available pQCD-based programs.
We include the nuclear shadowing modification of the parton distribution
functions and the in-medium radiative energy loss for heavy quarks, using
the mass-dependent BDMPS quenching weights.
Muons from W and Z leptonic decays, that dominate
the yield at high $p_{\rm t}$, can be used as a medium-blind
reference to observe the medium-induced suppression of beauty quarks.
\end{abstract}


\end{frontmatter}

% reset footnote counter (needed because of footnote in author list)
\setcounter{footnote}{0}

%%%%%%%%%%%%%%%%%%%%%%%%%%%%%%%%%%%%%%%%%%%%%%%%%%%%%%%%%%%%%%%%%%%%%%%%%%%
%             MAIN TEXT
%%%%%%%%%%%%%%%%%%%%%%%%%%%%%%%%%%%%%%%%%%%%%%%%%%%%%%%%%%%%%%%%%%%%%%%%%%%

\section{Introduction}
\label{intro}

Heavy quarks are regarded as effective
probes of the strongly-interacting medium
produced in ultra-relativistic heavy-ion collisions, since they are produced
in the initial hard-scattering processes
and they may subsequently interact with the
medium itself.
At the Relativistic Heavy Ion Collider (RHIC), a significant suppression
of the so called `non-photonic electrons', expected to be produced
in the semi-leptonic decay of charm and beauty hadrons, has been measured
in central Au--Au collisions at centre-of-mass energy $\sqrtsNN=200~\gev$
per nucleon--nucleon collision,
indicating a substantial energy loss of heavy quarks
in a strongly-interacting medium~\cite{PHENIX,STAR}.

In \mbox{Pb--Pb} collisions at the Large Hadron
Collider (LHC), the energy per nucleon--nucleon collision
will be of $5.5~\tev$, about 30 times larger than at
RHIC, opening up a new era for the study of
strongly-interacting matter at high energy
density (`QCD medium'). Heavy-quark
medium-induced energy loss will be one of the most captivating
topics to be addressed in this novel energy
domain~\cite{lhcpredictions}.
In hadron--hadron collisions at
LHC energies, muons are predominantly produced in semi-leptonic decays of
heavy-flavoured hadrons ---mostly beauty for muon $\pt\gsim 4~\gev/c$.
Thus, in heavy-ion collisions,
the muon $\pt$ distribution
is sensitive to b-quark energy loss effects.
In absence of nuclear effects,
the initial heavy-quark production yields are expected to scale
from proton--proton (pp)
to nucleus--nucleus collisions in a given centrality class,
according to the average number $\av{N_{\rm coll}}$
of inelastic nucleon--nucleon collisions
(binary scaling). Under this assumption, the in-medium energy loss
of heavy quarks would induce a suppression of the high-$\pt$
muon yield with respect to the binary-scaled yield measured in 
pp collisions.
The suppression can be
quantified as a reduction with respect to unity of the
nuclear modification factor:
\begin{equation}
\RAA(\pt,\eta) =
\frac{1}{\av{N_{\rm coll}}}\frac{\mathrm{d}^{2}N_{\rm AA}/\mathrm{d}\pt\mathrm{d}\eta}
{\mathrm{d}^{2}N_{\rm pp}/\mathrm{d}\pt\mathrm{d}\eta}\,.
\label{eq:RAA}
\end{equation}
However, initial-state effects,
like for example nuclear shadowing of the parton distribution
functions, could significantly reduce the initial production yields in
nucleus--nucleus collisions, thus making more difficult to relate
a reduction of $\RAA$ to b-quark energy loss.
At the LHC, muons from W and Z decays can provide an
intrinsic
calibration for the muon nuclear modification factor and a test of
the binary scaling assumption.
Due to the large amount of energy available at the LHC,
W and Z bosons will be produced with
significant cross sections 
in the hard parton--parton scatterings, and their initial
yields are expected to scale with $\av{N_{\rm coll}}$.
As we will show, the muons from the decays $\rm W\to\mu\nu_\mu$
and $\rm Z\to\mu\mu$
are predicted to dominate the muon $p_{\rm t}$ distribution for
$p_{\rm t}\gsim 30~$GeV/$c$.

Three experiments, ALICE~\cite{ALICEPPR1,ALICEPPR2},
ATLAS~\cite{ATLAS}, and CMS~\cite{CMS}, will measure the production of
muons in heavy-ion collisions, covering different acceptance regions.
The ALICE Muon Spectrometer covers the pseudo-rapidity range $2.5 <\eta< 4$
for $p > 4~\gev/c$ ($\pt\gsim 1~\gev/c$) . ATLAS
and CMS can measure muons at central pseudo-rapidity,
$|\eta| \lsim 2.5$, with a larger cutoff, $\pt>3$--$4~\gev/c$.
As we will quantify in section~\ref{sec:results}, the high-$\pt$
reach for the measurement of the inclusive muon production spectrum
is expected to extend well into the region where muons from W and Z
decays become dominant over muons from beauty decays.

In this work, we study the effect of heavy-quark energy
loss on the transverse momentum distribution of muons
in \mbox{Pb--Pb} collisions at $\sqrtsNN=5.5~\tev$ within
the acceptance of the LHC experiments.
The weak gauge bosons contributions (section~\ref{sec:WZ})
to the  muon
$p_{\rm t}$ and rapidity distributions
are obtained from the PYTHIA~\cite{PYTHIA} event generator, taking
into account the isospin content of the colliding nuclei
and normalizing the cross section to the values predicted by
the calculations in Refs.~\cite{FM04,Vog01}.
The heavy quark contributions (section~\ref{sec:HQ}) to
the  muon $\pt$ distribution are obtained from a NLO
perturbative QCD(pQCD)
calculation (MNR~\cite{HVQMNR}) supplemented with the
mass-dependent BDMPS quenching weights for radiative energy
loss~\cite{MassiveQuenching}, quark fragmentation \emph{\`a la}
Peterson~\cite{Peterson} and semi-muonic decay with the spectator
model~\cite{V-A}. In the calculation of heavy-quark energy loss the
decreased medium density at large rapidities is taken into account by assuming
the BDMPS transport coefficient $\hat q$
to scale as $\hat{q}(\eta)\propto
\d N_{\rm ch}/\d\eta$ (section~\ref{sec:Eloss}).
In section~\ref{sec:results} we present and discuss the resulting muon $\pt$
distribution in the 10\% most central \mbox{Pb--Pb} collisions,
the \mbox{Pb--Pb}-to-pp nuclear modification factor $\RAA(\pt)$ and the
central-to-peripheral nuclear modification factor $\RCP(\pt)$,
without and with the inclusion of heavy-quark energy loss.

\section{Muon $p_{\rm t}$ distribution in hadron--hadron collisions}
\label{sec:baseline}

The muon $\pt$ distributions in pp and \mbox{Pb--Pb} collisions at LHC energies
are calculated considering the semi-muonic decays of heavy-flavoured hadrons
and the muonic decay of W and Z bosons.
%Cold nuclear matter effects like shadowing and $k_t$ broadening have been included.
The procedure used to evaluate the production
differential cross section per nucleon--nucleon collision is described in the
following.

\subsection{$\rm W$ and $\rm Z$ decay muons}
\label{sec:WZ}

At the LHC, the c.m.s. energy is large enough to allow the production of
massive particles such as the W and Z bosons.
The $p_{\rm t}$ and rapidity distributions of W/Z and of their decay muons
are obtained from the PYTHIA event generator~\cite{PYTHIA},
that reproduces the measured $\pt$ distributions
in ${\rm p\overline p}$ collisions at the
Tevatron ($\sqrt{s}=1.8~\tev$)~\cite{MS99,BHP01}.
Since in hadron--hadron collisions,
at leading order, W and Z bosons are produced by
quark--anti-quark annihilation,
there could be a dependence of their production cross section on
the isospin of the input channel.
In nucleus--nucleus collisions this dependence can be
mimicked by a weighted cocktail
of proton--proton (pp), neutron--neutron (nn), proton--neutron (pn) and
neutron--proton (np) collisions. The
cross section per nucleon--nucleon binary collision can be expressed as
\begin{eqnarray}
\frac{\mathrm{d}^2 \sigma_{\scriptstyle\rm NN}}{\mathrm{d}p_{\rm t}\d y} & \approx&
\; \frac{Z^2}{A^2} \times
\frac{\mathrm{d}^2\sigma_{\rm pp}}{\mathrm{d}p_{\rm t} \d y} \; +
\; \frac{(A-Z)^{2}}{A^2} \times \frac{\mathrm{d}^2\sigma_{\rm nn}}{\mathrm{d}p_{\rm t} \mathrm{d}y} \; + \nonumber\\
& &\; \frac{Z \cdot (A-Z)}{A^2} \times \Bigg\{
\frac{\mathrm{d}^2\sigma_{\rm pn}}{\mathrm{d}p_{\rm t} \mathrm{d}y} \;
+ \; \frac{\mathrm{d}^2\sigma_{\rm np}}{\mathrm{d}p_{\rm t}
\mathrm{d}y} \Bigg\} \,,
\end{eqnarray}
where $A$ and $Z$ are the mass number
and the atomic number of the colliding nuclei.
CTEQ\,4L~\cite{CTEQ4M} parton distributions functions (PDFs) are
used, and nuclear shadowing is accounted for via the EKS98
parametrization~\cite{EKS98}.
The resulting $\pt$ distributions are normalized to the cross sections
obtained from Refs.~\cite{FM04,Vog01},
that is a cross section per nucleon--nucleon collision of $6.56$~($7.34$)~nb
for the W and $0.63$~($0.68$)~nb for the Z in \mbox{Pb--Pb} (pp) collisions at
$5.5$~TeV, including the muonic branching ratios (10.6\% for W and
3.4\% for Z~\cite{pdg}). %\cite{FM04,Vog01,Vog02}.
Notice that the Z production cross section is about ten times smaller
than that of the W~\cite{Zaida,Zaida_HQ}.
The uncertainty due to neglecting higher order corrections 
(next-to-next-to-leading order) was quantified 
as 1--2\% in Ref.~\cite{FM04} by varying the values of the factorization and 
renormalization scales. The uncertainty due to the errors on the 
PDFs was quantified as about 10\% in Ref.~\cite{ADMP04}.

\subsection{Heavy-quark decay muons}
\label{sec:HQ}

Within the pQCD collinear factorization framework, the expression
for the production cross section of heavy-flavoured hadrons in the collision of 
two hadrons $A$ and $B$ can be schematically written as:
\begin{eqnarray}
\frac{\mathrm{d}^2\sigma^{AB\to
h}_{\rm no\,medium}}{\mathrm{d}p_{\rm t}\d y}&=&\sum_{i,j}\int\mathrm{d}x_{i/A}~\mathrm{d}x_{j/B}~f_{i/A}(x_{i/A})~f_{j/B}(x_{j/B})\nonumber \times \\
&& \frac{\mathrm{d}^2\hat{\sigma}^{ij\to
{\rm Q}\overline{\rm Q} X}}{\mathrm{d}p_{\rm t,Q} \d y_{\rm Q}}\times \frac{D_{h/{\rm Q}}(z)}{z^{2}},
\label{eq:Factorization}
\end{eqnarray}
where $f_{i/A}(x_{i/A})$ and $f_{j/B}(x_{j/B})$ are the parton
distribution functions, the differential probabilities for the
partons $i$ and $j$ to carry momentum fractions $x_{i/A}$ and
$x_{j/B}$ of their respective nucleons.
$\hat{\sigma}^{ij}$ is the cross section of the partonic process
$ij\to {\rm Q}\overline{\rm Q} X$. The fragmentation function $D_{h/{\rm Q}}(z)$ is the
probability for the heavy quark Q to fragment into a hadron $h$ with
transverse momentum $p_{\rm t}=z\,p_{\rm t,Q}$.
 To simplify the notation, we have dropped the 
$\sqrt{s}$ dependence of $\hat{\sigma}^{ij}$ and 
the renormalization/factorization scale dependences of $f_{i/A(j/B)}$, 
$\hat{\sigma}^{ij}$ and $D_{h/{\rm Q}}$ 
(the squares of scales are normally of the order of
the momentum transfer $Q^2\sim p_{\rm t,Q}^2$ of the hard scattering).

We use the NLO pQCD calculation implemented in the HVQMNR program~\cite{HVQMNR}
to obtain the heavy-quark $\pt$--$y$ double-differential
cross sections, with the following
parameters values: for charm, $m_{\rm c}=1.2~\gev/c^2$ and factorization and
renormalization scales $\mu_{\rm F}=\mu_{\rm R}=2\mu_0$, where $\mu_0\equiv
\sqrt{m_{\rm Q}^2+(p_{\rm t,Q}^2 + p_{\rm t,\overline{Q}}^2)/2}$; for beauty,
$m_{\rm b}=4.75~\gev/c^2$ and $\mu_{\rm F}=\mu_{\rm R}=\mu_0$.
CTEQ\,4M~\cite{CTEQ4M} parton distribution functions are used, and nuclear
shadowing is taken into account with the EKS98
parametrization~\cite{EKS98}. For the b quark, 
the perturbative uncertainty was quantified, by varying the scales, 
in about 30\% for $\pt >30~\rm{GeV}/c$~\cite{heralhc}. 
Starting from the heavy-quark double-differential cross sections at NLO, 
we obtain the muon-level cross sections using the following Monte Carlo procedure.
We sample $\pt$ and $y$ of a c (or b) quark according to the shape of the NLO
cross section and fragment it to a hadron using the Peterson~\cite{Peterson}
fragmentation function % In this function we use the $\epsilon$ parameter values
% $\epsilon_{\rm c}=0.021$ and $\epsilon_{\rm b}=0.001$,
% extracted in a recent analysis of
% $\rm e^+ e^-$ data from LEP~\cite{heralhc}.
following the parameterization obtained in a recent analysis of
$\rm e^+ e^-$ data from LEP~\cite{heralhc}.
Finally, we decay the hadron into a muon according to the spectator
model~\cite{spectator}. In the spectator model, the heavy quark in a
meson is considered to be independent of the light quark and
is decayed as a free particle according to the $V-A$ weak
interaction~\cite{V-A}.
Thus, we assume the momentum of the hadron to be
entirely carried by the constituent heavy quark and we perform
the heavy-quark three-body decay, $\rm c\to s\mu\nu_\mu$ or
$\rm b\to c\mu\nu_\mu$, to obtain the muon transverse momentum and rapidity. 
The muon production cross sections per nucleon--nucleon collision
from charm (beauty) at $\sqrtsNN=5.5~\tev$
that we obtain are 0.415~mb (20~$\mu$b) in \mbox{Pb--Pb} collisions and 0.637~mb
(23~$\mu$b) in pp collisions. A charm (beauty) semi-muonic branching ratio
of 9.6\% (11.0\%)~\cite{pdg} has been considered.
We do not include the muons from the cascade decay $\rm b\to c\to\mu$,
because their yield is expected to become negligible with respect
to that of the direct muons from b, for $\pt$ larger than a few
$\gev/c$~\cite{notebe}.
The uncertainty on the beauty component, which dominates the heavy-flavour 
muon yields
in the $\pt$ range relevant to our study, can be quantified as about 30\%,
on the basis of a recent analysis~\cite{heralhc} 
that considered the perturbative uncertainty 
(variation of the factorization and renormalization scales), 
the uncertainty on the PDFs and the uncertainty 
on the fragmentation.

\section{Heavy-quark energy loss}
\label{sec:Eloss}

Now we will compute the muon $\pt$ distribution taking into account
the heavy-quark energy loss in the strongly-interacting medium that is
expected to be formed in central \mbox{Pb--Pb} collisions at LHC energies.
For modelling the energy loss of heavy quarks by medium-induced
gluon radiation,
we used the quenching weights in the multiple soft scattering
approximation, which were derived in Ref.~\cite{MassiveQuenching}
in the framework of the BDMPS formalism~\cite{BDMPS}. 
Schematically, energy loss
is introduced by modifying Eq.~(\ref{eq:Factorization}) to:
\begin{eqnarray}
\frac{\mathrm{d}^2\sigma^{AB\to
h}_{\rm medium}}{\mathrm{d}p_{\rm t}\d y}&&=\sum_{i,j}\int\mathrm{d}x_{i/A}~\mathrm{d}x_{j/B}~\mathrm{d}\Delta
E \times f_{i/A}(x_{i/A})~f_{j/B}(x_{j/B}) \times \nonumber \\
&& \frac{\mathrm{d}^2\hat{\sigma}^{ij\to
{\rm Q\overline Q} X}(p_{\rm t,Q}+\Delta E)}{\mathrm{d}p_{\rm t,Q} \d y_{\rm Q}}\times
P(\Delta E,\hat q,L,m_{\rm Q}/E)\times\frac{D_{h/\rm Q}(z)}{z^{2}},
\label{eq:Factorization_eloss}
\end{eqnarray}
where $E$ is the heavy-quark energy and $\Delta E$ is the
radiated energy. The quenching weight, represented by $P(\Delta E, \hat q, L, m_{\rm Q}/E)$, is the probability for a heavy quark with
mass $m_{\rm Q}$ and energy $E$ to lose
an energy $\Delta E$ while propagating over a path length $L$
inside a medium with transport coefficient $\hat q$. The latter is defined
as $\av{k_{\rm t}^2}/\lambda$, the average
transverse momentum, $k_{\rm t}$,
squared transferred from the medium to the parton
per unit mean free path $\lambda$. It is expected to be proportional
to the volume density $\d N^{\rm g}/\d V$ of gluons
in the medium (thus, to its energy
density) and to the typical momentum transfer per scattering.
%The average parton energy loss can be approximately related to
%$\hat q$ and $L$, as
%\begin{equation}
%\av{\Delta E}\propto \alpha_{\rm s}C_{\rm R}\hat q L^2\,,
%\end{equation}
%where $\alpha_{\rm s}$ is the strong coupling constant (we use
%the queching weights
%obtained with $\alpha_{\rm s}=0.3$~\cite{MassiveQuenching})
%and $C_{\rm R}$ is the Casimir coupling factor (3 for gluons
%and 4/3 for quarks).

We calculate the energy loss $\Delta E$
following the Monte Carlo approach introduced in Ref.~\cite{PQM}
for light quarks and gluons,
and adapted for heavy quarks in Refs.~\cite{MassiveQuenching,eleRHIC}.
We start by sampling the heavy-quark kinematics, $\pt$ and $y$, according to
the NLO double-differential cross section. Then,
we sample the parton production point in
the transverse plane ${\rm (x,y)}$
according to the Glauber-model~\cite{glauber} density
$\rho_{\rm coll}{\rm (x,y)}$ of binary
collisions, and we sample the azimuthal parton propagation direction.
We calculate the path length $L$
and the value of $\hat q$, which is a mean value of the local
time-averaged ($\overline{\phantom{a}}$) transport coefficient 
$\overline{\hat q}{\rm (x,y)}$ along the path of the
parton. 
We do not include the expansion of the medium in the longitudinal and 
transverse directions. However, it has been shown in Ref.~\cite{expansion} that,
numerically, the effects of a 
time-dependent medium on parton energy loss can be accounted for by an
equivalent static medium, specified in terms of the time-averaged 
transport coefficient $\overline{\hat q}$. This is confirmed 
by recent works (see e.g. Ref.~\cite{renk}),
showing that the inclusion of the medium evolution by a 
full hydrodynamical simulation does not significantly change or improve,
with respect to the assumption of a static medium,
the results for the suppression of the high-$\pt$ $\RAA$ in central 
\mbox{Au--Au}
collisions at RHIC.
We assume that our approach for the estimation of $L$
is valid for rapidities belonging to the ``central plateau'' of the
charged particle distribution~\cite{Bjorken}, 
which is expected to be as large as $|y|<4.5$ at
LHC energies~\cite{lhcpredictions}. 
We now sample a $\Delta E$ value, according to the quenching
weight $P$, and we modify the quark kinematics
to ($p_{\rm t}^{\prime}=p_{\rm t}-\Delta E$, $y^\prime=y$).
As in Ref.~\cite{MassiveQuenching}, 
heavy quarks that lose all their energy ($\Delta E>\pt$)
are redistributed
according to the thermal distribution
$\d N/ \d m_{\rm t}\propto m_{\rm t}\,\exp{\left(-{m_{\rm t}/T}\right)}$ 
with $T=0.3~\gev$. The muons from the decay of these thermalized quarks
populate the region $\pt\lsim 2~\gev/c$, 
outside the range of interest for the present study.
We assume the heavy-quark rapidity to stay constant during the process
of energy loss, since heavy quarks co-move with the longitudinally-expanding
medium and any modification of the initial heavy-quark rapidity should
remain small. Finally, we apply fragmentation and decay as described in 
section~\ref{sec:HQ}.

In Ref.~\cite{PQM} the local $\hat q$ transverse profile at central
rapidity ($y=0$) is
assumed to be proportional to the density of binary collisions. Since we
want to study muon production in a broad rapidity range ($|y|\lsim 4$),
we introduce a dependence of the local $\hat q$ on the pseudo-rapidity $\eta$
in order to account for the reduced medium density in the forward direction.
Namely, we assume the transport coefficient $\hat{q}$
to scale as a function of  pseudo-rapidity
according to the gluon pseudo-rapidity
density of the medium $\d N^{\rm g}/\d\eta$, this choice being justified
by the fact that $\d N^{\rm g}/\d V$ scales in $\eta$ according
to $\d N^{\rm g}/\d\eta$.
We assume the pseudo-rapidity density of charged particles and the
pseudo-rapidity density of gluons to
have the same dependence on $\eta$ and we write~\cite{HengtongQM06}:
\begin{equation}
\overline{\hat{q}}({\rm x,y},\eta)=\kappa\cdot\rho_{\rm coll}({\rm x,y})\cdot\Bigg[\frac{\mathrm{d}N_{\rm ch}}{\mathrm{d}\eta}(\eta)
\Bigg/\frac{\mathrm{d}N_{\rm ch}}{\mathrm{d}\eta}(0)\Bigg]\,,
\label{eq:QhatScale}
\end{equation}
where $\kappa$ is a
constant that sets the scale of $\hat q$. We use two values of $\kappa$
that were estimated for central \mbox{Pb--Pb} collisions at $\sqrtsNN=5.5~\tev$
by scaling the values extracted from an analysis of the light-flavour
hadrons suppression at RHIC~\cite{PQM,fragility}. The corresponding values for
the parton-averaged ($\av{\phantom{q}}$) and time-averaged 
($\overline{\phantom{q}}$) 
transport coefficient\footnote{Hereafter indicated as $\hat q$, 
for simplicity.} 
are $\av{\overline{\hat q}}=25$ and $100~\gev^2/\fm$.
For the pseudo-rapidity dependence we use the pseudo-rapidity
distribution of charged particles predicted for \mbox{Pb--Pb} collisions
at the LHC in Ref.~\cite{CGC}.
The effect of this dependence on the suppression of muons
from heavy-quark decays is expected to be small,
because $\mathrm{d}N_{\rm ch}/\mathrm{d}\eta$ is predicted
to have only a modest variation in the range $|\eta|<4$
(for illustration, for $\eta =3$ it would be reduced by about $15\%$
with respect to $\eta=0$).

Before presenting our results,
we point out that, in addition to radiative energy loss, 
there are other possible medium-induced effects 
(e.g.: collisional energy loss of heavy quarks~\cite{wicks}, 
in-medium hadronization and dissociation of heavy-flavour hadrons~\cite{adil}, 
formation of bound states and hadronization of heavy quarks via 
coalescence~\cite{vanhees}) 
that could noticeably affect the muon spectrum 
for $p_{\rm t}\lsim 10~\gev/c$. 
These effects have not been considered in the present 
study. And one has to keep in mind that
all current jet quenching models do not describe the suppression of single non-photonic
electron $\pt$ spectra at RHIC equally well as they do describe the suppression of pion spectra.
It is with this caveat that one should view the predictions for LHC based on these models.

\section{Results and discussion}
\label{sec:results}

\begin{figure*}[!t]
  \begin{center}
    \resizebox{\textwidth}{!}{
    \includegraphics{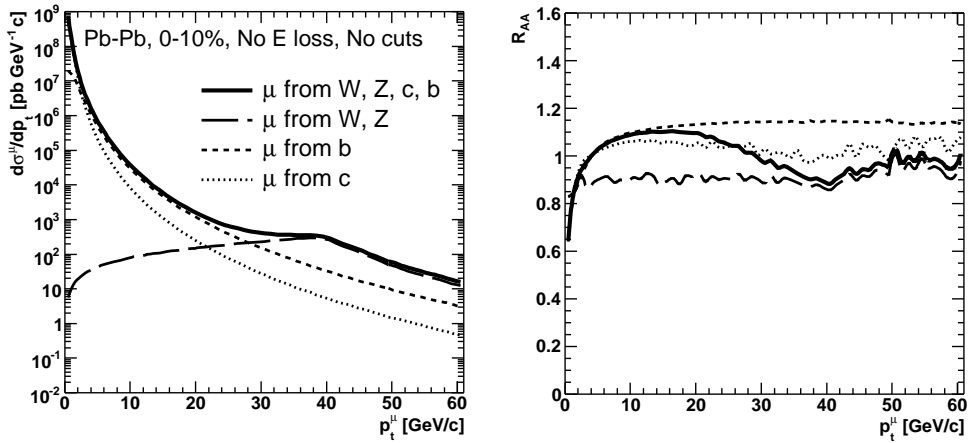}}
    \caption{Muons in
             central ($0$--$10\%$) \mbox{Pb--Pb} collisions
             at $\sqrtsNN=5.5$~TeV. Heavy-quark energy loss is not
             included and no acceptance cuts are applied.
         Left-hand panel:
             $\pt$-differential cross section normalized to one binary
         nucleon--nucleon collision.
             Right-hand panel: nuclear modification factor $\RAA$ with respect
             to pp collisions.
             \label{fig:1}}
  \end{center}
\end{figure*}

We start by presenting, in the left-hand panel of Fig.~\ref{fig:1}, the
muon production
cross-section as a function of transverse momentum in the 10\% most
central \mbox{Pb--Pb} collisions at
$\sqrtsNN=5.5$~TeV, when only nuclear shadowing is included.
The contributions from charm, beauty and weak gauge
bosons are shown separately. Muons from charged pion and kaon decays
and from Drell-Yan processes ($\rm q\overline q\to\mu^+\mu^-$)
are ignored here, because both are expected to be negligible in
the transverse momentum range $5\lsim\pt\lsim 60~\gev/c$~\cite{ALICEPPR1,CMS}.
Due to their large masses, W and Z bosons are mainly produced with small 
transverse momentum, $\pt\ll m_{\rm W,Z}$. Therefore,
 the decay muons have typically $\pt\sim m_{\rm W,Z}/2$. The latest 
qualitatively explains the shape of the $\pt$ distribution of muons 
from the decay of the W and Z, which ``peaks'' at $\pt\approx 40~\gev/c$.
Among the contributions that we compute,
muons from charm decays are predominant in the low-$\pt$ range,
2--$4~\gev/c$. In the range 4--$30~\gev/c$ beauty
decays prevail, and at larger $p_{\rm t}$ the W decays represent the largest
contribution to the  muon spectra.
We may note that these values for the ``crossing points'' in $\pt$
are somewhat dependent on the perturbative
uncertainties of NLO calculations
(choice of the fragmentation and renormalization scales)
and on other systematics in the fragmentation and decay kinematics.
For example, a variation of $\pm 30\%$ of yield of the beauty component
(see section~\ref{sec:HQ})
implies a shift of the crossing point by approximately $\pm 2~\gev/c$.
As discussed in Refs.~\cite{Zaida,Zaida_HQ},
the experimental measurement of the ratio of positive-to-negative muons
could help to determine these crossing points. 
For illustration of the expected significance for high-$\pt$ muons, 
we quote the production yield of muons from W decay in 
minimum-bias \mbox{Pb--Pb} collisions at $\sqrtsNN=5.5~\tev$ for 
an integrated luminosity of $5\times10^{32}~{\rm cm^{-2}}$ (the expected integrated luminosity to be 
collected by each of the three experiments 
within one month of data-taking).
In this case, about $7.5\times 10^4$ muons from W decays will be produced 
in the range $30<\pt<50~\gev/c$. Out of them, about $6.0\times 10^4$ 
will be produced in $|\eta | < 2.5$ and $0.7\times 10^4$ in $2.5 < \eta < 4.0$.

In view of exploring the final-state effects (heavy-quark
energy loss) on
the muon spectrum, we first focus on the influence of the
initial-state effects, i.e. nuclear shadowing, by analysing the
nuclear modification factor $\RAA$ as a function of transverse
momentum
(right-hand panel of Fig.~\ref{fig:1}).
The short-dashed and dotted lines represent the heavy-quark decays
contributions; at low $p_{\rm t}$ ($2$--$4~\gev/c$) they probe the small
$x$ range of the gluon PDF $g(x)$, where, according to the
EKS98 parametrization~\cite{EKS98}, we have shadowing: the PDF in the Pb
nucleus is suppressed with respect to the PDF in the free proton
($C_{\rm shad}(x,Q^2)=g^{\rm Pb}(x,Q^2)/g^{\rm p}(x,Q^2)<1$). Thus, $\RAA<1$.
For larger muon $p_{\rm t}$ ($\gsim 10~\gev/c$)
a higher $x$ range is probed, and we begin to explore
the anti-shadowing region ($C_{\rm shad}>1$), so $\RAA>1$ as is
observed in Fig.~\ref{fig:1}.
% In comparison, the nuclear shadowing makes a larger influence on muons from charm than from beauty decays.
For illustration, in table~\ref{tab:1} we report a qualitative
estimation of the probed $(x_1,x_2)$ values and of the corresponding
EKS98 shadowing factors $C_{\rm shad}=C_{\rm shad}(x_1,Q^2)\times C_{\rm shad}(x_2,Q^2)$ 
for heavy quarks and W/Z bosons
in \mbox{Pb--Pb} collisions at $5.5$~TeV,
as a function of the decay-muon rapidity and transverse momentum 
($\overline{C_{\rm shad}}$ is the mean value of $C_{\rm shad}$).
The $x$ values were obtained using the PYTHIA event generator for the
leading order processes $\rm gg\to Q\overline Q$ and $\rm q\overline q\to W/Z$.
For these two processes, we have, qualitatively, $s_{\scriptscriptstyle\rm NN}x_1 x_2=Q^2\approx 4\,(p_{\rm t,\mu}^2+m^2_{\rm Q})$ and
$s_{\scriptscriptstyle\rm NN}x_1 x_2=Q^2\approx m_{\rm W\,(Z)}^2\approx 4\,p_{\rm t,\mu}^2$, respectively; and, for both, $x_1\approx 2\,p_{\rm t,\mu}\exp(+y_{\mu})/\sqrtsNN$, $x_2\approx 2\,p_{\rm t,\mu}\exp(-y_{\mu})/\sqrtsNN$.
% The nuclear shadowing influence on weak boson decays (long-dashed line in Fig.~\ref{fig:1}) fluctuate around its mean value, $0.9$.
Weak gauge boson decays (long-dashed line in Fig.~\ref{fig:1})
probe the quarks nuclear shadowing, which fluctuates around its mean
value of $0.9$.
The overall nuclear modification factor of muons
(solid line in Fig.~\ref{fig:1}) increases rapidly with $p_{\rm t}$ up
to a value of about $1.1$ and then decreases to about $0.9$.
% Testing weak bosons at larger $p_{t}$ also means increasing the probed quarks $x$-range. %, while the 'large-$x$' value is in the shadowing humpback plateau region.
% At mid-rapidity, to augment $p_{t}$ is to increase both quarks $x$-range while keeping close values(reaching $x\aplt0.1$), that is to increase the value of the probed nuclear shadowing, from shadowing to anti-shadowing.
% At large rapidity, to augment $p_{t}$ is to increase one $x$ value and decrease the other, the large-$x$ value being approximately $x\apgt 0.1$, that is going from shadowing to anti-shadowing and to shadowing again.
% Those effects can not be observed in the whole rapidity interval(right-hand panel of Fig.~\ref{fig:1}) due to the competition of the whole, but they can be observed when testing a restricted acceptance window(lower plots of Fig.~\ref{fig:3}).
% (at mid-rapidity) and
% When testing weak bosons in a restricted acceptance window the probed nuclear shadowing results from the competition of a)
In order to explore the uncertainties due to the limited knowledge of the nuclear PDFs in the kinematic region relevant to our study, we considered two 
other parametrizations of the nuclear parton distribution: nDS~\cite{nDS} and HKN07~\cite{HKN}.
The values of the corresponding shadowing factors are reported in 
table~\ref{tab:1}. % We note that the shadowing factor uncertainty for heavy quarks at high $\pt$ is 10\% and for weak gauge boson is less than 5\%. For heavy quarks at low $\pt$, it becomes as large as 25\% for charm and 10\% for beauty. 
We note that the shadowing factor uncertainty for heavy quarks at high $\pt$ is 7\% and for weak gauge boson is less than 5\%. For heavy quarks at low $\pt$, it becomes as large as 20\% for charm and 10\% for beauty.  
Note that the study of proton--nucleus collisions at RHIC energies 
has been absolutely necessary to disentangle cold and hot nuclear matter 
effects. 
Cold nuclear matter effects at LHC energies remain relatively unknown. 
Therefore, p--Pb runs at the LHC will be needed in order to fully understand
the nuclear modification factors measured in Pb--Pb.

\begin{table}[!htbp]
  \begin{center}
      \caption{Qualitative estimation of Bjorken-$x$ values and shadowing factors (according to EKS98~\cite{EKS98}, nDS(NLO)~\cite{nDS} and HKN07(NLO)~\cite{HKN}) for heavy quarks and W/Z
      bosons produced in \mbox{Pb--Pb} collisions at $\sqrtsNN=5.5~\tev$ as a function of the decay-muon $y$ and $\pt$ (in $\gev/c$). Details in the text.
      \label{tab:1}
    }
      \begin{tabular}{cccccccccccc} \hline  \noalign{\smallskip}
        &  $y$ & $p_{\rm t}$ & $x_{1}$ & $x_{2}$  & $C_{\rm shad}^{EKS}$  & $\overline{ C_{\rm shad}^{EKS} }$  & $C_{\rm shad}^{nDS}$  & $\overline{C_{\rm shad}^{nDS} } $ & $C_{\rm shad}^{HKN}$ & $\overline{ C_{\rm shad}^{HKN} }$  \\  \noalign{\smallskip}\hline \noalign{\smallskip}
    \multirow{4}{*}{$\rm c$} 
    & 0 & 0 &  \multicolumn{2}{c}{$4\cdot 10^{-4}$} & 0.50 & \multirow{4}{*}{0.65} &0.81& \multirow{4}{*}{0.92}&0.74&\multirow{4}{*}{0.87}\\
        & 0 & 30 &   \multicolumn{2}{c}{ $1\cdot 10^{-2}$} & 1.10 &  & 1.09& &0.93&  \\
        & 3 & 0 &  $9\cdot 10^{-3}$ &$2\cdot 10^{-5}$ & 0.57 &  & 0.83&  &0.74&\\
        & 3 & 30 & $2\cdot 10^{-1}$ & $5\cdot 10^{-4}$ & 0.91 &  &0.96 & & 1.09&\\
        \noalign{\smallskip}\hline \noalign{\smallskip}
    \multirow{4}{*}{$\rm b$} 
    & 0 & 0  &  \multicolumn{2}{c}{$2\cdot 10^{-3}$} & 0.77 & \multirow{4}{*}{0.85} & 0.92&\multirow{4}{*}{0.95} & 0.83&\multirow{4}{*}{0.93} \\
        & 0 & 30 &   \multicolumn{2}{c}{$1\cdot 10^{-2}$} & 1.10 &  &1.00 & & 0.93& \\
        & 3 & 0 &  $3\cdot 10^{-2}$ & $8\cdot 10^{-5}$  & 0.85 &  &0.93 & & 0.85&\\
        & 3 & 30 &  $2\cdot 10^{-1}$ & $5\cdot 10^{-4}$ & 0.91 &  & 0.96& &1.09& \\
         \noalign{\smallskip}\hline \noalign{\smallskip}
    \multirow{2}{*}{W} 
    & 0 & all &  \multicolumn{2}{c}{$1\cdot 10^{-2}$} & 0.89 & \multirow{2}{*}{0.89} &0.88& \multirow{2}{*}{0.83} & 0.83&\multirow{2}{*}{0.84} \\
        & 3 & all &  $3\cdot 10^{-1}$ & $7\cdot 10^{-4}$  & 0.76 &  &0.77 & &0.85& \\
        \noalign{\smallskip}\hline \noalign{\smallskip}
    \multirow{2}{*}{Z} 
    & 0 & all & \multicolumn{2}{c}{$2\cdot 10^{-2}$} & 0.90 & \multirow{2}{*}{0.93} &0.89& \multirow{2}{*}{0.83} & 0.84&\multirow{2}{*}{0.83}\\
        & 3 & all &  $3\cdot 10^{-1}$ & $8\cdot 10^{-4}$ &  0.77 & & 0.76& & 0.83&
        \\ \noalign{\smallskip} \hline  \noalign{\smallskip}
      \end{tabular}
  \end{center}
\end{table}

\begin{figure*}[htb]
  \begin{center}
    \resizebox{0.8\textwidth}{!}{
      \includegraphics{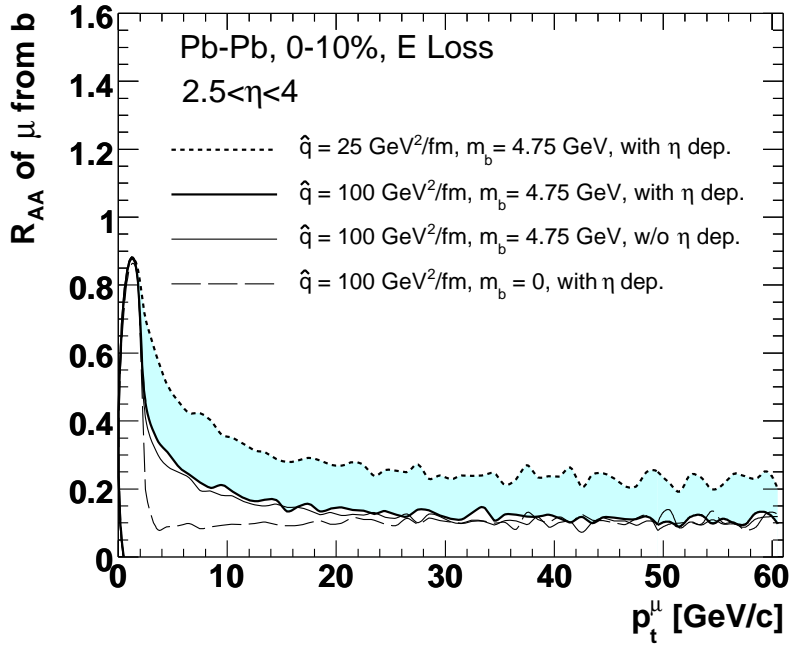}}
    \caption{Nuclear modification factor of muons from beauty decays with/without mass effect in the energy loss, and with/without pseudo-rapidity dependence of
$\hat q$
             ($\mathrm{d}N/\mathrm{d}\eta$ dependence), in the central (0--10\%) \mbox{Pb--Pb} collisions at $\sqrtsNN=5.5~\tev$ in the pseudo-rapidity range
$2.5<\eta<4.0$.
             \label{fig:2}
           }
  \end{center}
\end{figure*}

\begin{figure*}[!htb]
  \begin{minipage}[c]{0.48\linewidth}
    \centering\resizebox{\textwidth}{!}{
      \includegraphics{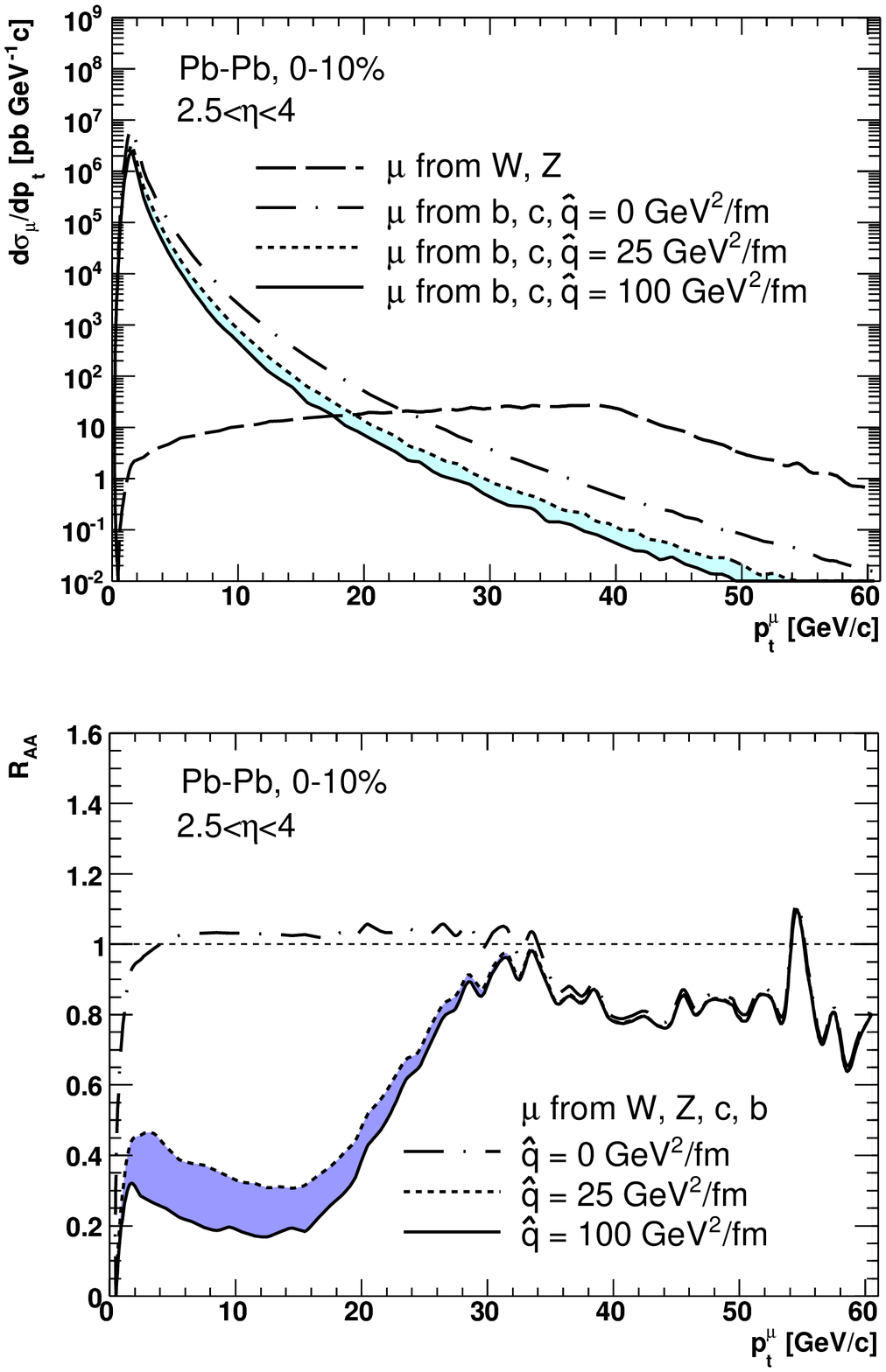}}
  \end{minipage}
  \begin{minipage}[c]{0.48\linewidth}
    \centering\resizebox{\textwidth}{!}{
      \includegraphics{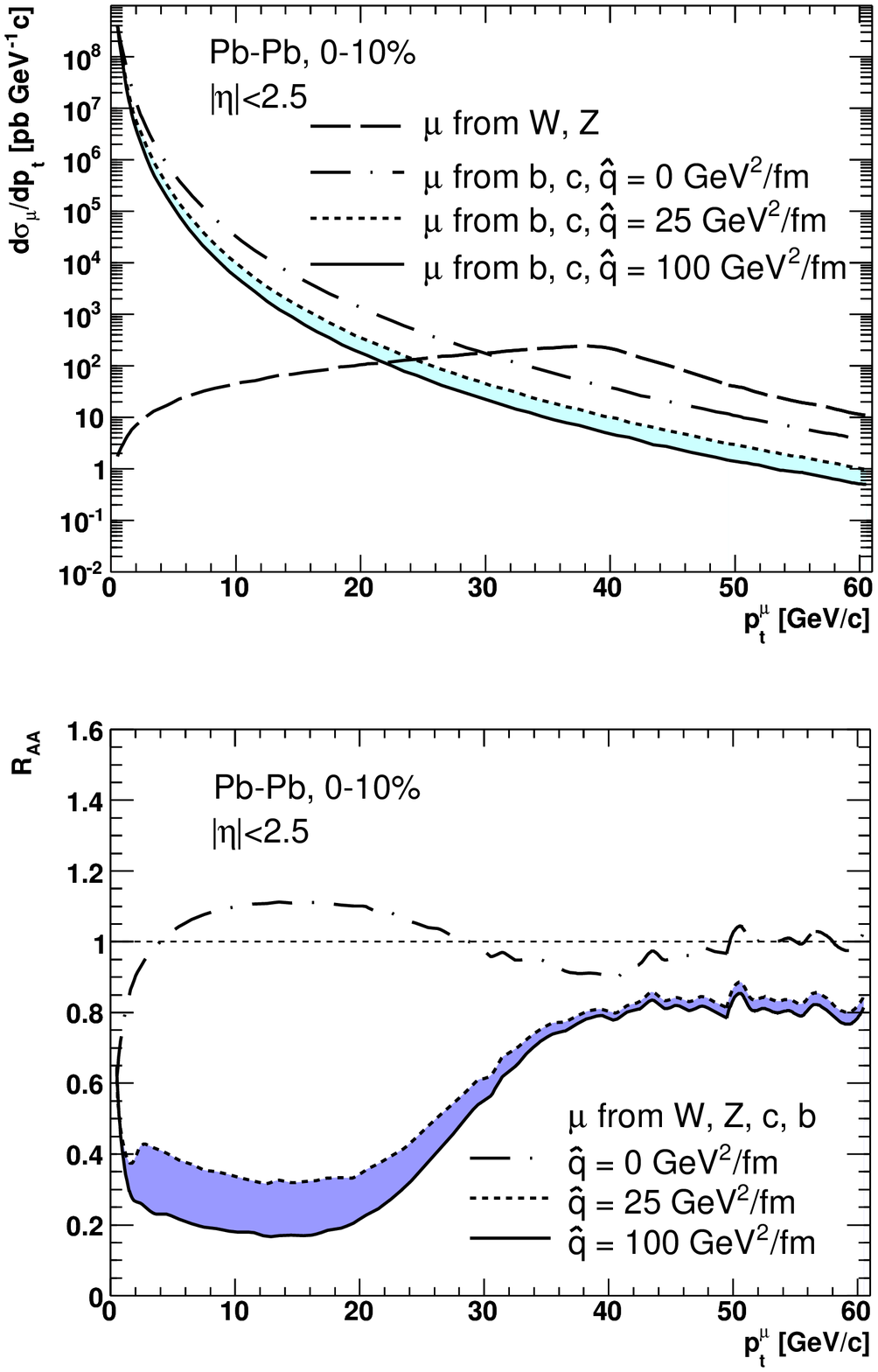}}
  \end{minipage}
     \caption{Differential cross section per nucleon--nucleon
       collision and nuclear modification factor of muons
              from W, Z, c and b decays in central (0--10\%) \mbox{Pb--Pb} collisions at $\sqrtsNN=5.5$~TeV.  for $2.5<\eta<4$.
             The left-hand panel shows the results for $2.5<\eta<4$, % within the acceptance of ALICE muon spectrometer
      the right-hand panel the results for $|\eta|< 2.5$.% within the acceptance of CMS/ATLAS.
              \label{fig:3}
            }
\end{figure*}

\begin{figure*}[!htb]
  \begin{minipage}[c]{0.48\linewidth}
    \centering\resizebox{\textwidth}{!}{
      \includegraphics{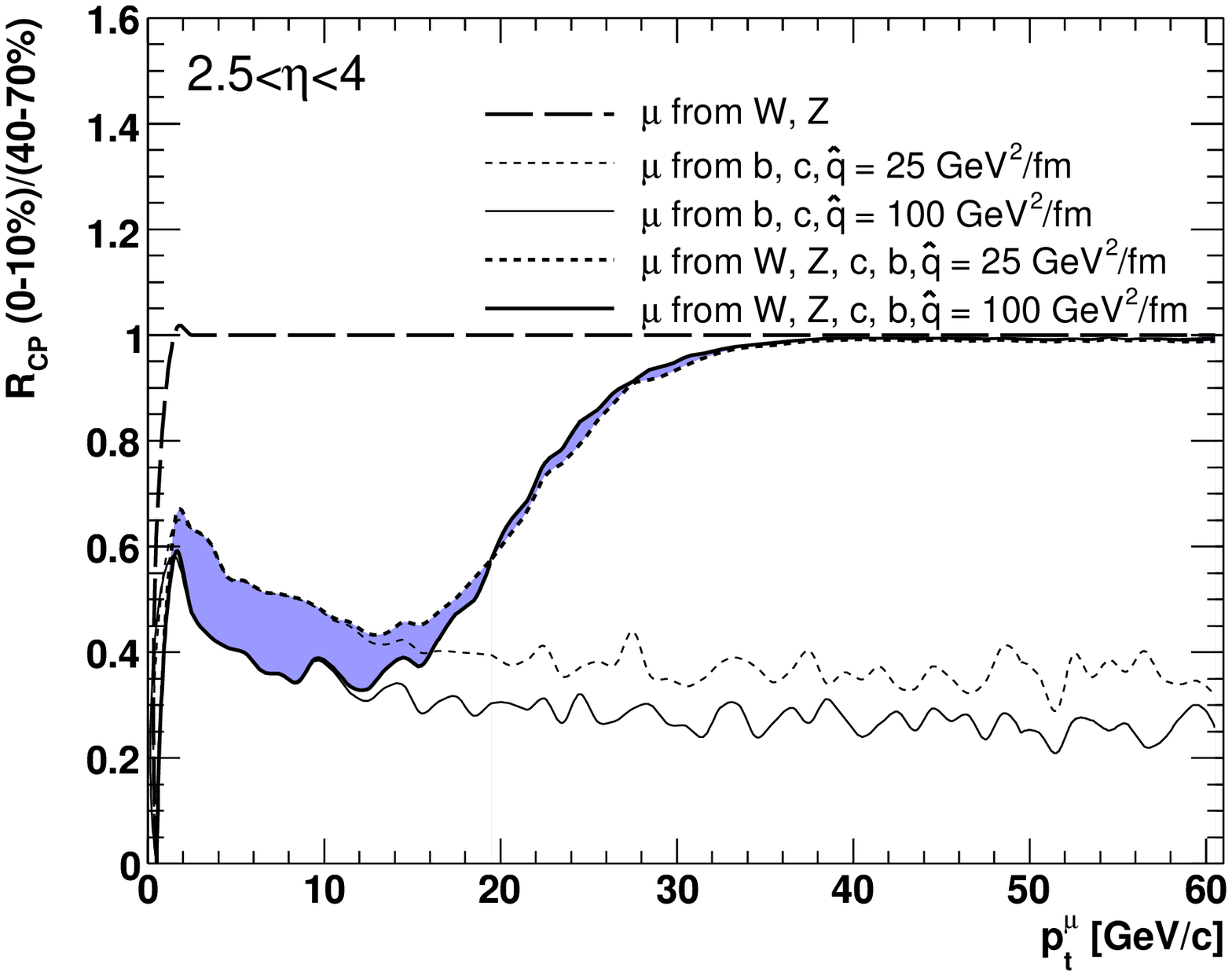}}
  \end{minipage}
  \begin{minipage}[c]{0.48\linewidth}
    \centering\resizebox{\textwidth}{!}{
      \includegraphics{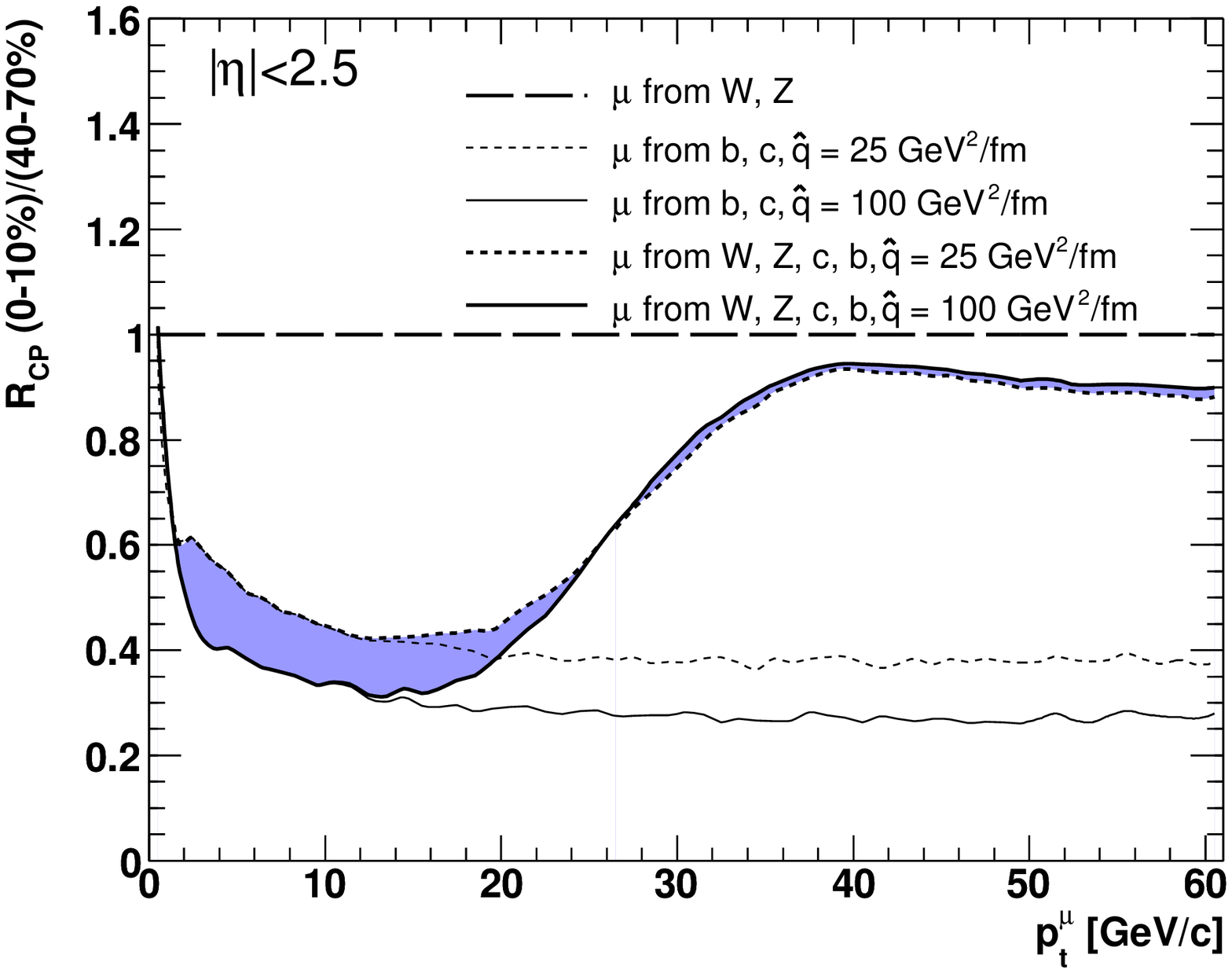}}
  \end{minipage}
     \caption{Central (0--10\%) to peripheral (40--70\%) nuclear modification factors of muons from W, Z, c and b decays in \mbox{Pb--Pb} collisions at $\sqrtsNN=5.5$~TeV.
      The left-hand panel shows the results for $2.5<\eta<4$, % within the acceptance of ALICE muon spectrometer
      the right-hand panel the results for $|\eta|< 2.5$. % within the acceptance of CMS/ATLAS.
      \label{fig:4}
    }
\end{figure*}

We now include in the calculation the in-medium energy loss for
heavy quarks. We start by considering the nuclear modification
factor $\RAA(\pt)$ of muons from beauty decays in central
(0--10\%) \mbox{Pb--Pb} collisions, in order to study the effects of the
b quark mass and of the dependence of the transport coefficient
$\hat q$ on $\eta$ according to $\d N_{\rm ch}/\d\eta$. The latter
is relevant only in the large pseudo-rapidity range. The result is
shown in Fig.~\ref{fig:2}. The shaded band represents our baseline
result for the $\hat q$ range 25--100~$\gev^2/\fm$, with mass
effect ($m_{\rm b}=4.75~\gev$) and with pseudorapidity dependence
of $\hat q$. The suppression obtained at high $\pt$, where $\RAA$
becomes independent of $\pt$, is about a factor 5 (10) for $\hat
q=25~(100)~\gev^2/\fm$. The suppression for electrons from beauty
decays was calculated within the same framework (except for the
treatment of hadronization and decay) in
Ref.~\cite{MassiveQuenching}, for central rapidity and $\pt\lsim
15~\gev/c$. For the same $\pt$ range, we obtain similar values for
$\RAA$. With reference to Fig.~\ref{fig:2}, by comparing the thick
solid line ($m_{\rm b}=4.75~\gev$) and the long-dash line ($m_{\rm
b}=0$), we notice that the quark mass effect in parton energy loss
increases $\RAA$ by up to a factor of three for $\pt\sim
5~\gev/c$, and that some effect persists even at $15~\gev/c$. When
going to $p_{\rm t}\gsim 20~\gev/c$, the quark mass dependence
becomes negligible since the quark mass becomes negligible with
respect to the momentum. By comparing the thick and thin solid
lines with $\hat{q}=100~\gev^2/\fm$, which are with and without
$\eta$ dependence of $\hat q$ respectively, we conclude that this
dependence has basically no effect on the amount of suppression in
the range $2.5<\eta<4$.

Figure~\ref{fig:3} shows the muon
$\pt$-differential cross section and the nuclear
modification factor for muons from W, Z, c and b decays in
central (0--10\%) \mbox{Pb--Pb} collisions at
$\sqrtsNN=5.5$~TeV % within the acceptance of ALICE
% muon spectrometer and CMS/ATLAS($|\eta|<2.5$)
in the two pseudo-rapidity domains. The results with transport
coefficient values $\hat q=0$, 25 and 100~GeV$^2$/fm are
reported, where $\hat{q}=0$ corresponds to the
case of no energy loss. With reference to the upper panels of
Fig~\ref{fig:3}, note that, since muons from W and Z decays are unaffected by
the energy loss, the crossing point in transverse momentum
of the distributions of
b-quark and W-boson decay muons shifts down by $\approx 5~\gev/c$ at large
rapidities and by $\approx 7~\gev/c$ at mid-rapidity, when energy loss is
included.
% From the shift of the crossing
% point, which maybe observed by the experiments, one may get the
% influence of medium-induced effects on heavy quarks theoretically.
Concerning the muon $\RAA(\pt)$ (lower panels of Fig.~\ref{fig:3}),
at large pseudo-rapidities (on the left),
with heavy-quark energy loss, the overall
muon yield is suppressed by about a factor of 2--5 in the range
$2<p_{\rm t}< 20~\gev/c$,
where the beauty contribution dominates.
For higher $\pt$,
$\RAA$ increases rapidly in the $20 \aplt p_{\rm t} \aplt 30~\gev/c$
range and flattens at around 0.8 above
$\approx 30~\gev/c$. The
$\hat q$-independence of the $\RAA$ of overall muons at large
$p_{\rm t}$ is due to the fact that the W/Z boson
contribution to the yield becomes dominant (see upper panels of
Fig.~\ref{fig:3}). % So this $\pt$ domain %only
% reflects the nuclear shadowing effect on the partons and then it
% can be useful to study the cold nuclear effects.
At mid-rapidity the behaviour is similar (lower-right panel).
The small difference of the $\RAA$ shape at different rapidities is due
to the different proportion of heavy-quark and W/Z boson decays.
% The nuclear
% modification factor without energy loss in the central rapidity
% firstly increases rapidly and then decreases to about 0.90 when at
% about 1.10. It is due to that W/Z decay muons dominates the large
% $\pt$ range and the their shadowing factor is about 0.9, and heavy
% quarks dominates the small $\pt$ range and their shadowing factor
% could be large than 1.(see that in Tab. 1)
For the same reason the nuclear modification factor without energy loss
(dot-dashed lines in the lower panels in Fig.~\ref{fig:3}) differs
slightly from that obtained without acceptance cuts
(solid line in Fig.~\ref{fig:1}).
% Testing weak bosons at larger $p_{t}$ also means increasing the probed quarks $x$-range. %, while the 'large-$x$' value is in the shadowing humpback plateau region.
% At mid-rapidity, to augment $p_{t}$ is to increase both quarks $x$-range while keeping close values(reaching $x\aplt0.1$), that is to increase the value of the probed nuclear shadowing, from shadowing to anti-shadowing.
% At large rapidity, to augment $p_{t}$ is to increase one $x$ value and decrease the other, the large-$x$ value being approximately $x\apgt 0.1$, that is going from shadowing to anti-shadowing and to shadowing again.
% Those effects can not be observed in the whole rapidity interval(right-hand panel of Fig.~\ref{fig:1}) due to the competition of the whole, but they can be observed when testing a restricted acceptance window(lower plots of Fig.~\ref{fig:3}).

Besides the \mbox{Pb--Pb}-to-pp nuclear modification factor
$\RAA$, also the central-to-peripheral nuclear modification factor $\RCP$
will provide information on the medium-induced suppression of b quarks.
$\RCP$ is defined as:
\begin{equation}
\RCP(p_{\rm t}) = \frac{\langle
N^{AA}_{\mathrm{coll}}\rangle^{\mathrm{P}}}{\langle
N^{AA}_{\mathrm{coll}} \rangle^{\mathrm{C}}}
~\frac{\mathrm{d}^2{N_{AA}^{\mathrm{C}}}/\mathrm{d}p_{\rm t}\mathrm{d}y}
      {\mathrm{d}^2{N_{AA}^{\mathrm{P}}}/\mathrm{d}p_{\rm t}\mathrm{d}y}\,,
\label{eq:RCP}
\end{equation}
where the index $\mathrm{C\,(P)}$ stands for
central (peripheral) collisions.
From the experimental point of view, the $\RCP$ measurement will be
more straight-forward than the $\RAA$ measurement, for the following two 
reasons.
1) The measurements in pp and in \mbox{Pb--Pb} will be affected by 
different systematic errors (especially for the cross section normalization),
which will add up in the $\RAA$ uncertainty. 2) pp collisions at the LHC will
have different c.m.s. energy (14~TeV) with respect to \mbox{Pb--Pb} (5.5~TeV), 
therefore the muon spectra measured in pp will have to be extrapolated 
from 14~TeV to 5.5~TeV with the guidance of perturbative QCD calculations, 
introducing an additional systematic error 
of the order of 10\% on $\RAA$~\cite{ALICEPPR2}.

In our calculation, the initial-state effects are assumed
to be the same in central and peripheral
collisions ---namely, we do not include an impact parameter dependence for
shadowing--- thus, they cancel out in the central-to-peripheral ratio.
As a consequence,
the $\RCP$ of muons from weak gauge boson decays is equal to one.
The central (0--10\%) to peripheral (40--70\%) ratios are shown in
Fig.~\ref{fig:4}.
In central (0--10\%) collisions the
yield might be reduced with respect to peripheral
collisions (40--70\%)
by a factor 2--3 in the $p_{\rm t}$ range from about $2~\gev/c$
to about $13~\gev/c$, where the b-quark contribution dominates.
When going to larger $p_{\rm t}$,
the $\RCP$ of muons
increases fast and then flattens at around 0.8 at
mid-rapidity and 1.0 at forward rapidity. This difference at high $\pt$
between the two pseudo-rapidity regions is due to the different
relative abundances of the heavy-quarks and weak bosons components.
For the same reason,
the curves for $\hat{q}=25~\gev^2/\fm$
and $100~\gev^2/\fm$ cross each other at $\pt\approx 20~\gev/c$
at large pseudo-rapidity and at $\pt\approx 25~\gev/c$
at mid-rapidity.
We have checked that the uncertainties on the cross sections of muons from 
W decays and of muons from beauty decays (approximately 
10\% and 30\%, respectively,
as discussed in section~\ref{sec:baseline}) 
translate into a variation smaller than 5\%
of the $\RAA$ and $\RCP$ values for $\pt\gsim 35~\gev/c$, while they have no
effect at lower $\pt$.

\section{Conclusions}
\label{sec:summary}

The effect of heavy-quark energy loss on the differential cross section of
muons produced in \mbox{Pb--Pb} and pp collisions at LHC energies has been
investigated.
The most important contributions to the decay muon yield in the range $5<\pt<60~\gev/c$ %$2<\pt<60~\gev/c$ 
have been included: b (and c) quarks have been
computed using a NLO pQCD
calculation supplemented with the BDMPS mass-dependent
quenching weights; weak gauge bosons muonic decays
have been computed using the PYTHIA
event generator.
The heavy-quark mass-dependence of energy loss reduces the suppression of
 muon yields from beauty decays by about factor two for
$5\lsim\pt\lsim 15~\gev/c$.
To account for the decrease of the medium density at large pseudo-rapidity,
we assumed a decrease of the transport coefficient
proportionally to $\mathrm{d}N_{\rm ch}/\mathrm{d}\eta$,
and we found that the effect on the muon $\pt$ distribution is negligible,
especially at large transverse momentum.
We investigated the energy
loss effect by means of the \mbox{Pb--Pb}-to-pp and of the \mbox{Pb--Pb}
central-to-peripheral nuclear modification factors
in the acceptance of the LHC experiments: ATLAS, ALICE, and CMS.
In the $\pt$ interval below approximately $20~\gev/c$, where the beauty
component is dominant, $\RAA$ for 0--10\% central \mbox{Pb--Pb} collisions 
relative to
pp and $\RCP$ for 0--10\% relative to 40--70\% \mbox{Pb--Pb} collisions are 
found to
be about 0.2--0.4 and 0.3--0.5, respectively.
Then, we observe a steep rise in the beauty/W crossover interval
20--$40~\gev/c$, up to the values $\RAA\approx 0.8$ and $\RCP\approx 1$
in the interval above $40~\gev/c$, dominated by W/Z decay muons.
These muon nuclear modification factors could provide the first
experimental observation of the b quark medium-induced suppression
in \mbox{Pb--Pb} collisions at the LHC.
The presence of a medium-blind component (muons from W and Z decays)
that dominates the high-$\pt$ muon yield will allow an intrinsic
calibration of the medium-sensitive probe (heavy quarks), because it
will provide a handle on the strength of the initial-state effects
that may alter the hard-scattering cross sections in nucleus--nucleus
collisions at the unprecedented energies of the LHC.

\paragraph*{Acknowledgments.}
The authors, members of the ALICE Collaboration, would like to
thank their ALICE Colleagues for useful exchanges during the
accomplishment of the present work. In particular, we would like
to thank F.~Antinori, P.~Crochet, A.~Morsch and J.~Schukraft for
fruitful discussions, N.~Armesto and C.~A.~Salgado for their
helpful suggestions at the beginning of this work and the calculation of the HKN parameterized nPDFs, 
and to S.~Kumano and R.~Sassot for the calculation of the HKN and the nDS parameterized nPDFs, and to M. Mangano for the HVQMNR program implementation.
\\
This work is partly supported by
the NSFC (10575044 and 10635020), the Key Project of the Chinese Ministry of Education (306022 and IRT0624)
and
the France China Particle Physics Laboratory, FCPPL (CNRS/IN2P3 and Chinese Ministry of Education).

\end{document}